\documentclass[%
reprint,
superscriptaddress,
nofootinbib,
amsmath,amssymb,
jap
]{revtex4-1}

\usepackage[utf8]{inputenc}
\usepackage[T1]{fontenc}
\usepackage{dcolumn}% Align table columns on decimal point
\usepackage{bm}% bold math
\usepackage{makecell}
\usepackage{multirow}
\usepackage{xcolor}
\usepackage{graphicx}% Include figure files
\usepackage{textcomp, gensymb}
\usepackage{siunitx}
\usepackage{booktabs}
\usepackage{setspace}
\usepackage{xr}% import labels from other files
\usepackage{placeins}
\usepackage[caption=false]{subfig}
\usepackage[export]{adjustbox}

\usepackage{braket} % from OAH
    
\usepackage{natbib}
\setcitestyle{super}

\usepackage{outlines}
\usepackage{enumitem}
\setenumerate[1]{label=\Alph*.}
\setenumerate[2]{label=\roman*.}
\setenumerate[3]{label=\alph*.}
\newcommand{\bigO}{\mathcal{O}}

\newcommand{\angstrom}{\mbox{\normalfont\AA}}
\makeatletter
\let\cat@comma@active\@empty
\newcommand*{\addFileDependency}[1]{% argument=file name and extension
  \typeout{(#1)}% latexmk will find this if $recorder=0 (however, in that case, it will ignore #1 if it is a .aux or .pdf file etc and it exists! if it doesn't exist, it will appear in the list of dependents regardless)
  \@addtofilelist{#1}% if you want it to appear in \listfiles, not really necessary and latexmk doesn't use this
  \IfFileExists{#1}{}{\typeout{No file #1.}}% latexmk will find this message if #1 doesn't exist (yet)
}
\makeatother

\newcommand*{\myexternaldocument}[1]{%
    \externaldocument{#1}%
    \addFileDependency{#1.tex}%
    \addFileDependency{#1.aux}%
}
\myexternaldocument{supporting-information}

\begin{document}

\title{Static Subspace Approximation for Random Phase Approximation Correlation Energies: Implementation and Performance}

\author{Daniel Weinberg}
\affiliation{Applied Mathematics \& Computational Research Division, Lawrence Berkeley National Laboratory,  Berkeley, CA, USA}

\author{Olivia A. Hull}
\affiliation{Materials, Chemical, and Computational Science Directorate, National Renewable Energy Laboratory, Golden, CO, USA}

\author{Jacob M. Clary}
\affiliation{Materials, Chemical, and Computational Science Directorate, National Renewable Energy Laboratory, Golden, CO, USA}

\author{Ravishankar Sundararaman}
\affiliation{Department of Materials Science and Engineering, Rensselaer Polytechnic Institute, Troy, NY, USA}

\author{Derek Vigil-Fowler$^\dag$}
\affiliation{Materials, Chemical, and Computational Science Directorate, National Renewable Energy Laboratory, Golden, CO, USA}

\author{Mauro Del Ben$^\dag$}
\affiliation{Applied Mathematics \& Computational Research Division, Lawrence Berkeley National Laboratory,  Berkeley, CA, USA}

\def\thefootnote{\dag}\footnotetext{Corresponding authors: MDB (mdelben@lbl.gov), D.V-F. (derek.vigil-fowler@nrel.gov)}\def\thefootnote{\arabic{footnote}}

\date{\today}

\begin{abstract}
\noindent Developing theoretical understanding of complex reactions and processes at interfaces  requires using methods that go beyond semilocal density functional theory to accurately describe the interactions between solvent, reactants and substrates. Methods based on many-body perturbation theory, such as the random phase approximation (RPA), have previously been limited due to their computational complexity. However, this is now a surmountable barrier due to the advances in computational power available, in particular through modern GPU-based supercomputers. In this work, we describe the implementation of RPA calculations within BerkeleyGW and show its favorable computational performance on large complex systems relevant for catalysis and electrochemistry applications. 
Our implementation builds off of the static subspace approximation which, by employing a compressed representation of the frequency dependent polarizability, enables the evaluation of the RPA correlation energy with significant acceleration and systematically controllable accuracy. 
We find that the  computational cost of calculating the RPA correlation energy  scales only linearly with system size for systems containing up to 50 thousand bands, and is expected to scale quadratically thereafter. We also show excellent strong scaling results across several supercomputers, demonstrating the performance and portability of this implementation. 
\end{abstract}

\maketitle

\section{Introduction}
%look at BGW paper
Electrocatalytic reactions undergird the key recent innovations and necessary future advances to decarbonize the world economy, yet there remains a patchwork of theoretical and experimental frameworks within which to understand these crucial reactions. Within the computational sphere, there has been significant progress in determining the correct reaction pathways, thermodynamics and kinetics following a grand-canonical density function  approach.~\cite{doi:10.1073/pnas.1713164114,C7CP08153G,10.1063/1.4978411,doi:10.1021/acs.jpclett.6b00358,doi:10.1021/jacs.8b10016}
Yet, while density-functional theory (DFT) has provided crucial insight into the microscopic structure of chemical processes, there remains a need for rigorous \textit{ab-initio} methods such as those based on many body perturbation theory to give chemical insight beyond the DFT level of theory.~\cite{doi:10.1021/jp002302t,Schimka2010,PhysRevB.75.085435}  

The application of many-body perturbative methods has generally been limited to smaller systems compared to DFT approaches due to their increased computational complexity. Of these methods, the random phase approximation (RPA) has the most favorable complexity of $O(N^4)$, but is still significantly more expensive than semilocal DFT calculations.~\cite{Ren2012} In particular for implementations in a plane-wave basis, this expense has been perceived as too computationally demanding due to the larger basis set size as campared to, for example, implementations using a localized basis. Therefore RPA's applications are not yet widely adopted and in most cases are restricted to bulk crystalline solids, small molecules, and small surface supercells of catalytic systems which enforces high coverages of single adsorbates.~\cite{PhysRevB.77.045136,PhysRevLett.103.056401,PhysRevLett.101.266106,PhysRevB.80.045402,Schimka2010} 
Approaches to reduce the computational cost in a plane-wave basis representation, for example by reducing the scaling of the calculation effort with respect to system size from quartic to cubic, have helped to push the limit of RPA to larger applications.~\cite{PhysRevB.90.054115,doi:10.1021/acs.jctc.3c00615,PhysRevB.101.205145,10.1063/1.4939841,doi:10.1021/acs.jctc.6b01235,10.1063/1.4921542,PhysRevB.109.035103,doi:10.1021/acs.jctc.3c01157,doi:10.1021/acs.jctc.2c00512}
Yet the increased complexity of the RPA approaches faces challenges and restrictions that prevents a more complete exploration of the complex interfacial systems of interest to the electrocatalysis community at the beyond-DFT level of theory. 
Furthermore, the RPA correlation energy is notoriously slow to converge with respect to the plane-wave basis set size and extrapolation techniques are necessary to obtain accurate energetic as required in the context of catalysis and electrochemistry.~\cite{PhysRevB.77.045136,PhysRevB.86.035111} The use of localized basis sets has shown success at reducing the cost of the evaluation of the RPA correlation energies by exploiting compact basis representations and sparsity.\cite{del_ben_electron_2013,doi:10.1021/acs.jctc.6b00840,PhysRevB.109.035103,doi:10.1021/acs.jctc.8b00177,doi:10.1021/acs.jctc.6b01235,LU2017187,10.1063/1.5090605} A plane-wave basis, however, provides for a systematic and system-agnostic method for extrapolation to the complete basis set limit, as the convergence is smooth and well understood. \cite{PhysRevB.77.045136,PhysRevB.81.115126}
This is a very appealing property especially for highly heterogeneous systems, where the basis set dependence can hinder the streamlining of the computational workflow for production calculations.

Recently, advances in computing power, especially the rise of GPU-based supercomputer architectures, have made RPA calculations on larger system models increasingly tractable.~\cite{DELBEN2015120,doi:10.1021/acs.jctc.3c01157,doi:10.1021/acs.jpclett.0c00320,doi:10.1021/acs.jctc.1c00494,10.1063/5.0144493,doi:10.1021/acs.jctc.2c00241} In this paper we present the implementation of the RPA correlation energy within the BerkeleyGW software package. The evaluation of the RPA correlation energy is sped up by using the static subspace approximation which employs a low-rank approximation from the plane-wave basis to a subspace basis within which to express the frequency dependent polarizability.~\cite{bgw_subspace} This method enables a reduction of the computational cost for the RPA correlation energy  proportional to the square of the ratio of the full basis to the subspace basis. This compression factor can also be tuned to for the appropriate trade off between approximation and computational expense.  
Furthermore, the memory requirements are reduced by an algorithmic strategy that divides the computation of the RPA correlation energy into blocks with variable number of valence bands (\texttt{NV-block} algorithm). In the limit, this algorithm reduces the cubic memory requirements to scale only quadratically with system size. This implementation, achieved with the portable OpenMP-target and OpenACC programming models, is thoughtfully benchmarked across various GPU-based architecture of modern high-performance computing (HPC) systems, achieving exceptional performance, scaling and portability. 

This implementation also incorporates partial occupations into the calculation, enabling the effects of finite temperature on the correlation energy to be studied at the RPA level. This advance is crucial to understanding the electrocatalytic systems of particular interest to this study. The ability to efficiently calculate highly accurate \textit{ab-initio} correlation energies for large catalytic  systems is a major step towards achieving an understanding of these systems that goes beyond the DFT level of theory.

\section{Method}

The RPA correlation energy is given by
\begin{equation}
    E_c^\text{RPA} = \sum_{\mathbf{q}} \int_0^\infty \dfrac{\text{d}\omega}{2\pi} \operatorname{Tr} \left[ \operatorname{ln}\left(1 -\chi^0(\mathbf{q},i\omega)v\right) + \chi^0(\mathbf{q},i\omega)v\right], \label{eqn:rpaE}
\end{equation}
where $\chi^0(\mathbf{q},i\omega)$ is the irreducible polarizability at point $\mathbf{q}$ in the Brillouin zone with imaginary frequency $i\omega$, and $v$ is the bare coulomb interaction.  The trace is taken over the basis for representing the polarizability, traditionally the reciprocal lattice vectors of the system.

Starting from the typical expression of the independent particle (or Kohn-Sham) response function given by Adler\cite{rpa1} and Wiser\cite{rpa2}, we have:

\begin{equation}
\begin{split}
    \chi^{0}_{\mathbf{G},\mathbf{G}'}(\mathbf{q}, \omega)= \frac{1}{V}\sum_{\mathbf{k}}\sum_{n}\sum_{n'}2g_{\mathbf{k}}(f_{n',\mathbf{k+q}}-f_{n,\mathbf{k}})\label{eq:chi0_definition}\\
    \times\frac{\braket{\psi_{n',\textbf{k}+\textbf{q}}|{e^{i(\textbf{q}+\textbf{G})\cdot \textbf{r}}}|{\psi_{n,\textbf{k}}}}\braket{\psi_{n,\textbf{k}}|{e^{-i(\textbf{q}+\textbf{G}')\cdot \textbf{r}'}}|{\psi_{n',\textbf{k}+\textbf{q}}}}}{\epsilon_{n',\textbf{k}+\textbf{q}} -\epsilon_{n,\textbf{k}} - i\omega} ,
\end{split}
\end{equation}
where $V$ is the volume of the real-space unit cell, $\mathbf{k}$ the crystal momentum vector or $\mathbf{k}$-point, $\epsilon_{n,\textbf{k}}$ the Kohn-Sham energy of band $n$ at $\mathbf{k}$-point $\textbf{k}$, $\textbf{G}$ the reciprocal lattice vector, $g_{k}$ the k-point weight,  $f_{n, \textbf{k}}$ the occupation factor of band $n$ at $\textbf{k}$, and $\textbf{q}$ and $\omega$ the specific point in the Brillouin zone and the frequency for which we are computing the response, respectively. 

Both $n$ and $n'$ run over all occupied and unoccupied bands. If both $n$ and $n'$ are occupied, then $f_{n, \textbf{k}} = f_{n', \textbf{k}+\textbf{q}} = 1$, so any term in the summation of this type is zero. Similarly, if $n$ and $n'$ are both unoccupied, then $f_{n, \textbf{k}} = f_{n', \textbf{k}+\textbf{q}} = 0$, so these types of terms are also zero. In the case of the system having partial occupations, we have four total types of non-zero terms in the summation with respect to the bands: (occupied, empty), (occupied, partially occupied), (partially occupied, empty), and (partially occupied, partially occupied). 

Letting

\begin{equation}
M_{nn'\textbf{k}}^\textbf{G}(\textbf{q})
\equiv
\braket{\psi_{n\textbf{k}+\textbf{q}}|{e^{i(\textbf{q}+\textbf{G})\cdot \textbf{r}}}|{\psi_{n'\textbf{k}}}} ,\label{eq:pwMat}
\end{equation}
and grouping the terms in the summation into these types of pairings, we can write:

\begin{equation}
\begin{split}
    &\chi^{0}_{\mathbf{G},\mathbf{G}'}(\mathbf{q}, \omega)= \frac{1}{V}\sum_{\mathbf{k}}2g_{\mathbf{k}}  \label{eq:chi0_partial_occs_sum_1}\\
    \times \left[\vphantom{\sum_{n}^{\text{occ}}}\right.&\sum_{n}^{\text{occ}}\sum_{n'}^{\text{emp}} \frac{(f_{{n'},\mathbf{k+q}}-f_{{n},\mathbf{k}})} {\epsilon_{{n'},\textbf{k}+\textbf{q}} -\epsilon_{{n},\textbf{k}} - i\omega}[M_{nn'\textbf{k}}^\textbf{G}(\textbf{q})]^* M_{nn'\textbf{k}}^{\textbf{G}'}(\textbf{q})\\
    +&\sum_{n}^{\text{occ}}\sum_{n'}^{\text{partial}} \frac{(f_{n',\mathbf{k+q}}-f_{n,\mathbf{k}})}{\epsilon_{n',\textbf{k}+\textbf{q}} -\epsilon_{n,\textbf{k}} - i\omega}  [M_{nn'\textbf{k}}^\textbf{G}(\textbf{q})]^* M_{nn'\textbf{k}}^{\textbf{G}'}(\textbf{q})\\
    +&\sum_{n}^\text{{partial}}\sum_{n'}^\text{{emp}} \frac{(f_{n',\mathbf{k+q}}-f_{n,\mathbf{k}})} {\epsilon_{n',\textbf{k}+\textbf{q}} -\epsilon_{n,\textbf{k}} - i\omega} [M_{nn'\textbf{k}}^\textbf{G}(\textbf{q})]^* M_{nn'\textbf{k}}^{\textbf{G}'}(\textbf{q})\\
    +&\sum_{n}^\text{{partial}}\sum_{n'}^\text{{partial}} \frac{(f_{n',\mathbf{k+q}}-f_{n,\mathbf{k}})}{\epsilon_{n',\textbf{k}+\textbf{q}} -\epsilon_{n,\textbf{k}} - i\omega} [M_{nn'\textbf{k}}^\textbf{G}(\textbf{q})]^* M_{nn'\textbf{k}}^{\textbf{G}'}(\textbf{q})\left.\vphantom{\sum_{n}^\text{{partial}}}\right] .\\
\end{split}
\end{equation}
We can re-write this equation with a change of indices, letting band index $v$ (for valence) include occupied and partially occupied bands, and band index $c$ (for conduction) include partially occupied and empty bands. Further defining 
\begin{equation}
    \Delta_{vc\mathbf{k}}(\mathbf{q},\omega) = \frac{2g_{\mathbf{k}}(f_{c,\mathbf{k+q}}-f_{v,\mathbf{k}})} 
    {\epsilon_{c,\textbf{k}+\textbf{q}}-\epsilon_{v,\textbf{k}} - i\omega} ,
\end{equation}
this yields:

\begin{equation}
    \label{eq:chi0_partial_occs}
    \chi^{0}_{\mathbf{G},\mathbf{G}'}(\mathbf{q}, \omega)= 
    \frac{1}{V}\sum_{\mathbf{k}}\sum_{v}^\text{val}\sum_{c}^\text{cond}
    [M^\mathbf{G}_{vc\mathbf{k}}(\mathbf{q})]^* \Delta_{vc\mathbf{k}}(\mathbf{q},\omega) M^{\mathbf{G}'}_{vc\mathbf{k}}(\mathbf{q}).
\end{equation}
Note that if there are no partial occupations, then the occupation factor term disappears, which yields exactly the form traditionally implemented in BerkeleyGW for the treatment of semiconductors and insulators.~\cite{bgw1}

The frequency dependence of the polarizability is calculated using the static subspace approximation~\cite{bgw_subspace}. In this approximation the zero-frequency polarizabilty for a given $\mathbf{q}$-point, $\chi^{0}(\mathbf{q}, \omega=0)$, is calculated using equation \ref{eq:chi0_partial_occs}, which is then used to define a subspace within which to calculate the non-zero frequencies~\cite{west1,west2,west3,west4,west5}. The projection matrix onto this subspace is the $N_G \times N_b$ matrix $\mathbf{C}_s$ consisting of the $N_b$  eigenvectors of $\mathbf{\chi}^0$ with the greatest absolute eigenvalues. This matrix is used to project the plane-wave matrix elements into the subspace basis. This projection is written as
\begin{equation}
    \mathbf{M}_s(\mathbf{q}) = \mathbf{M}(\mathbf{q})\mathbf{C}_s .
\end{equation}
Accordingly, the $N_b\times N_b$ subspace representation of the polarizability can be calculated simply as
\begin{equation}
    \mathbf{\chi}^0_s = \mathbf{M}_s^\dagger\mathbf{\Delta}\mathbf{M}_s .
\end{equation}
%no diagonalization needed. LU trick. Tr{ln} = ln{det}

The final RPA correlation energy is then given by the integral in equation \ref{eqn:rpaE} which is carried out within the subspace basis. The use of the subspace approximation leads to an acceleration by the square of the ratio of the subspace basis to the full plane-wave basis.

\section{Implementation and GPU Optimization}

\begin{figure}
    \centering
    \includegraphics[trim={0 0 11.5cm 0cm },clip, width=\columnwidth] {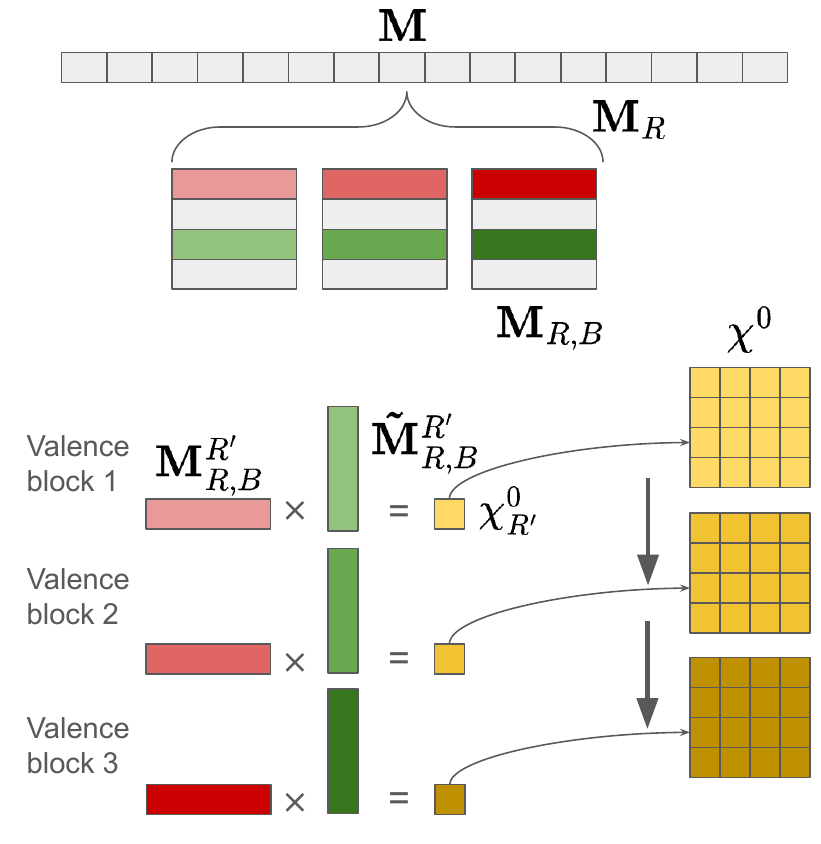}
    \caption{Diagrammatic representation of the valence block algorithm applied to the \texttt{CHI\_SUM} kernel using a 16 rank parallelization. The distributed matrix $M_{vc\mathbf{k}}^\mathbf{G}$ is shown with $\mathbf{G}$ indexed by the rows, and the combined $(v,c,\mathbf{k})$ index as the columns. The resulting $\chi^0$ matrix (shown in yellow) is accumulated over the valence blocks.}
    \label{fig:nvBlock}
\end{figure}

The RPA correlation energy within the static subspace approximation described here is implemented in the BerkeleyGW software package optimized for massively parallel multi-node and multi-GPU computations of excited state electronic properties of materials. BerkeleyGW employs methods and algorithms designed to perform at scale and capable of fully saturating leadership class HPC architectures~\cite{pseudobands2,DelBen2020_GB}. 
These modern GPU-based HPC systems can provide an  acceleration by orders of magnitude of intensive computational kernels. This, however, led to the exposure of previously overlooked bottlenecks on multi-core architectures which motivated a redesign of the core implementations to continue advancing the methods. 
What is required is a sophisticated, multi-layered parallelization scheme that is conscious of balancing  available memory, host to GPU offload, and communication resources. 
Furthermore, in order to achieve true performance portability across various GPU architectures, the directive based OpenACC and OpenMP-target programming models have been used to realize the implementation. This choice simplifies the maintenance, readability and development of GPU kernels while keeping the complexity of explicit interfaces limited to vendor specific library APIs.

The calculation of RPA energies in BerkeleyGW is handled by two main computational kernels. The \texttt{MTXEL} kernel calculates the plane-wave matrix elements $M_{cv\mathbf{k}}^\mathbf{G}$ in equation \ref{eq:pwMat}. These plane-wave matrix elements are then used in the \texttt{CHI\_SUM} kernel to calculate the frequency dependent polarizabilty matrix according to equation \ref{eq:chi0_partial_occs}.

Within the \texttt{MTXEL} kernel the natural parallelization is over the set of valence-conduction pairs. At the highest level of parallelization, the conduction bands are distributed over MPI ranks. This makes sense because converged RPA/GW calculations require the consideration of an extremely large number of unoccupied (conduction) states and thus these can be divided over a large number of processors. When extremely large systems with many valence states are divided over a great number of MPI ranks, it can be the case that each rank only is responsible for a small number of conduction states compared to the number of valence states owned. In this scenario an additional layer of parallelism is implemented wherein the valence states are now divided as well. In this case a given set of conduction states will be assigned to a number of ranks, each of which has a different set of valence states, called a valence pool. This division is designed such that each rank should generally be responsible for approximately the same number conduction states and valence states.  

We will denote the portion of the plane-wave matrix elements $\mathbf{M}$ on MPI rank $R$ as 
\begin{equation}
    \mathbf{M}_{R} = \left [ M_{cv\mathbf{k}}^\mathbf{G} : c \in R, v\in \text{Pool}(R)\right ],
\end{equation}
which includes all conduction bands $c$ associated with rank $R$ and all valence bands $v$ associated with the valence pool $\text{Pool}(R)$.  
The plane-wave matrix elements for each rank are computed efficiently on the GPU through a batched FFT algorithm. The computational time on each rank should scale as $\bigO \left( N_R^{-1} N_k N_v N_c N_{G,\psi} \log N_{G,\psi} \right)$. Where $N_R$, $N_k$, $N_v$, and $N_c$ are the number of MPI ranks, $\mathbf{k}$-points, valence bands, and conduction bands, respectively, and $N_{G,\psi}$ is the basis set size for the wave-functions.   

The  parallelization scheme for the \texttt{CHI\_SUM} kernel must take these distributed conduction-valence pairs, perform the summation in equation \ref{eq:chi0_partial_occs} in this distributed manner and then re-distribute the resulting polarizability matrix into a 2-D block cyclic data layout over the $\mathbf{G}$ vectors. This re-distribution is achieved by splitting the matrix multiplication into the sub-matrices needed to accumulate the portion of $\mathbf{\chi}^0$ owned by a particular rank. We will denote the block of $\mathbf{\chi}^0$ owned by rank $R'$ as 
\begin{equation}
    \mathbf{\chi}^0_{R'} = \left[\chi^0_{\mathbf{G},\mathbf{G'}} : (\mathbf{G},\mathbf{G'}) \in \operatorname{Grid}(R')\right] ,
\end{equation}
where $\operatorname{Grid}(R')$ is the set of $(\mathbf{G},\mathbf{G'})$ pairs for which $\chi^0_{\mathbf{G},\mathbf{G'}}$ is stored in memory on rank $R'$. 
This block of the matrix can be represented as a matrix multiplication between sub-matrices of $\mathbf{M}$ and the diagonal $\mathbf{\Delta}$ matrix. The "row sub-matrix" of the part of $\mathbf{M}$ stored on rank $R$ needed to calculate the block of $\mathbf{\chi}^0$ to be stored on rank $R'$ will be called
\begin{equation}
    \mathbf{M}^{R'}_R = \left[ M_{cv\mathbf{k}}^\mathbf{G} \in  \mathbf{M}_{R} : \exists \mathbf{G}' \text{ where } (\mathbf{G},\mathbf{G'}) \in \operatorname{Grid}(R')\right],
\end{equation}
and the "column sub-matrix" for that block is
\begin{equation}
    \tilde{\mathbf{M}}^{R'}_R = \left[ M_{cv\mathbf{k}}^{\mathbf{G}} \in  \mathbf{M}_{R}: \exists \mathbf{G}' \text{ where } (\mathbf{G}',\mathbf{G}) \in \operatorname{Grid}(R')\right]^\dagger,
\end{equation}
such that
\begin{equation}
    \mathbf{\chi}^0_{R'} = \sum_R\mathbf{M}^{R'}_R\mathbf{\Delta}_R\tilde{\mathbf{M}}^{R'}_R, \label{eq:rankchisum}
\end{equation}
where $\mathbf{\Delta}_R$ is the restriction of $\mathbf{\Delta}$ to the appropriate $c,v,\mathbf{k}$ indices. These row and column sub-matrices are shown in diagrammatically in red and green respectively in Figure \ref{fig:nvBlock}, with the resulting $\mathbf{\chi}^0_{R'}$ sub-matrix shown as the small yellow square. The $\mathbf{\Delta}_R$ matrix is omitted for visual clarity. For the zero-frequency polarizability, the computational time on each rank should scale as $\bigO \left( N_R^{-1} N_k N_v N_c N_G^2\right)$, where $N_G$ is the size of the $\mathbf{G}$-space basis. For the non-zero frequencies, the computation should scale as $\bigO \left( N_R^{-1} N_k N_v N_c N_b^2\right)$, where $N_b$ is  the size of the subspace basis. The sum over ranks $R$ is implemented via a previously described non-blocking cyclic communication scheme that hides the communication overhead by transferring data while the matrix multiplication is being performed on the GPU.~\cite{pseudobands2} 

The size of the full $\mathbf{M}$ matrix grows as $\bigO \left( N_k N_v N_c N_G\right)$. This matrix can easily grow to several hundred gigabytes or even several terabytes of data. If the entire matrix needs to be loaded into memory at once, this creates a lower bound on the amount aggregate available memory necessary to perform a calculation. Performing the matrix multiplication in equation \ref{eq:rankchisum} on GPUs presents additional challenges. Even if there is sufficient aggregate memory available, the on-chip memory of the GPU is generally not sufficient to hold $\mathbf{M}_R$. In previous implementations, to ensure there was sufficient GPU memory, the row and column sub-matrices $\mathbf{M}_R^{R'}$, $\tilde{\mathbf{M}}_R^{R'}$, and $\mathbf{\Delta}$ were prepared on the CPU and then transferred to the GPU for multiplication. This implementation was highly inefficient as it required a significant amount of CPU-GPU communication which often took longer than the computation time of the matrix multiplication on the GPU. This communication should scale as $\bigO(N_G N_v N_c N_k N_R^{-\frac{1}{2}})$. This dependence on the number of ranks $N_R$ is undesirable as for large numbers of ranks the computational kernels will scale inversely with the number of ranks, but the communication will only scale as the inverse square root, causing the communication cost to dominate. 

\begin{figure}
    \centering
    \includegraphics[width=\columnwidth]{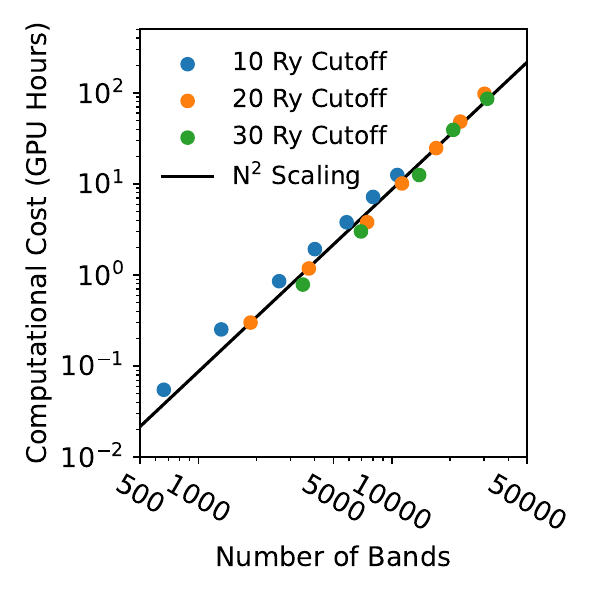}
    \caption{Total computational cost in Perlmutter GPU hours for calculating one $\mathbf{q}$-point in RPA calculations with 10 Ry, (blue) 20 Ry (orange) and 30 Ry (green) screened cutoff as a function of the number of bands in the calculation. The $N^2$ phenomenological scaling is shown to guide the eye.}
    \label{fig:sizePerf}
\end{figure}
\begin{figure}
    \centering
    \includegraphics[width=\columnwidth]{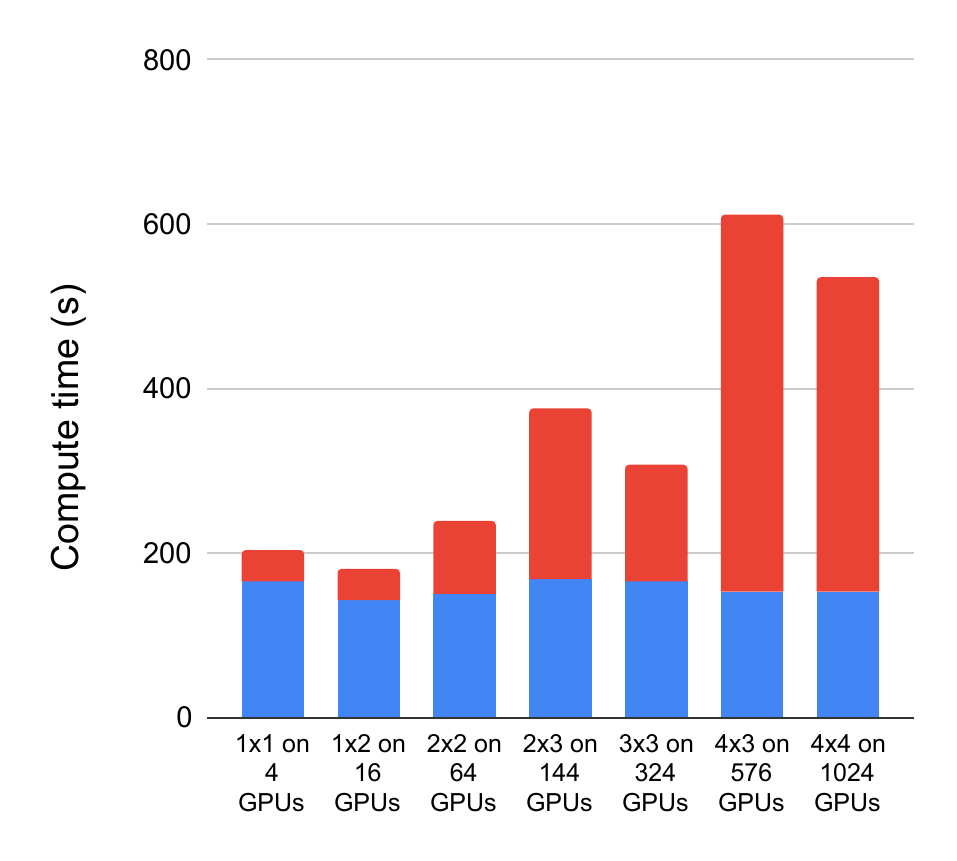}
    \caption{Weak scaling of the \texttt{MTXEL} (blue) and \texttt{CHU\_SUM} (red) kernels for $N^2$ scaling on Perlmutter using a 20 Rydberg screened cutoff.}
    \label{fig:weakScale}
\end{figure}

In order to circumvent this, an algorithmic strategy has been devised that enables to block the computation over a variable number of valence bands' batches (\texttt{NV-block} algorithm), where the number is determined by the maximum memory available per MPI rank (host or device).
Within the \texttt{NV-block} algorithm $\mathbf{M}_R$ is further subdivided into blocks small enough to fit into the GPU memory. These blocks are formed by dividing the valence states into sufficiently small groups. Each of these valence blocks are generated by \texttt{MTXEL} independently, significantly lowering the overall aggregate memory requirement. In fact, with this algorithm the limiting memory requirement is the size of the wave-functions on input, rather than any of the intermediate calculations. Figure \ref{fig:nvBlock} shows $\mathbf{M}_R$ divided into three such blocks. We will call these blocks
\begin{equation}
    \mathbf{M}_{R,B} = \left [ M_{cv\mathbf{k}}^\mathbf{G} \in \mathbf{M}_{R} : v\in \text{Block}(B,R)\right ],
\end{equation}
where $\text{Block}(B,R)$ is the set of valence states in block $B$ on rank $R$. Each block is transferred to the GPU only once and the equivalently defined block row and column sub-matrices, $\mathbf{M}^{R'}_{R,B}$, and $\tilde{\mathbf{M}}^{R'}_{R,B}$, are prepared on the GPU. The partial sum of $\mathbf{\chi}^0_{R'}$ after each block is accumulated over the blocks to give the complete sum. The summation of the polarizability, rewritten to highlight the computational division, becomes 
\begin{align}
    \mathbf{\chi}^0_{R'} &= \sum_\text{B}^{\text{blocks}} \sum_\text{R}^{\text{ranks}} \mathbf{M}^{R'}_{R,B}\mathbf{\Delta}_{R,B}\tilde{\mathbf{M}}^{R'}_{R,B}, 
\end{align}
where again $\mathbf{\Delta}_{R,B}$ is the appropriate restriction of $\mathbf{\Delta}_R$. This valence block algorithm, aside from shifting a substantial amount of computation from CPU to GPU, significantly alters the communication between CPU and GPU.  In the limiting case the communication is reduced by a factor of $2\sqrt{N_R}$ bringing the scaling of the communication to $\bigO(N_G N_v N_c N_k N_R^{-1})$. This means that the communication will now follow the same inverse dependence on the number of ranks as the computation, and will not come to dominate performance for large numbers of ranks.

It is worth noting that, within the static subspace approximation, this loop over blocks is performed twice. The initial loop is used to generate only the zero-frequency polarizability and generate the subspace basis. The second loop calculates the polarizability for all the remaining non-zero frequencies within the subspace basis. For this second pass, all the distribution remains the same, except the subspace basis instead of the $\mathbf{G}$ basis is used. As the goal of the valence block algorithm is to avoid ever storing the entirety of $\mathbf{M}$, these two stages require that $\texttt{MTXEL}$ recalculate the $\mathbf{M}$ matrix during the non-zero frequency part. This is not necessary in the case where the problem is small enough or sufficiently parallelized such that only a single valence block is needed. In that case $\texttt{MTXEL}$ is only run once as the entirety of $\mathbf{M}$ is already being stored. 

To finally calculate the correlation energy, the integral and trace in equation \ref{eqn:rpaE} must be calculated. While the trace is often calculated in the $\mathbf{G}$ basis, here we calculate it within the subspace basis. Additionally, rather than computing the logarithm of $1 -\chi^0v$, we use the identity $\operatorname{Tr}\left\{\mathrm{ln}(\mathbf{A})\right\}=\mathrm{ln}(\operatorname{det}(\mathbf{A}))$ and compute the determinant from an LU decomposition of the dielectric matrix: $\epsilon = 1 -\chi^0v$.\cite{DELBEN2015120} This method is preferred as the LU decomposition is computationally more efficient than matrix diagonalization and it allows the easy calculation of $\epsilon^{-1}$ which can be used in subsequent GW calculations. 

% \begin{align}
%     \chi^0_{\mathbf{G}\mathbf{G}'}(\mathbf{q},\omega) &= \sum_\text{B}^{\text{blocks}} \sum_\text{R}^{\text{ranks}} \nonumber\\
%     &\sum_{v\in \text{B}} \sum_{c\in \text{R}} \sum_\mathbf{k} [M^\mathbf{G}_{vc\mathbf{k}}(\mathbf{q})]^* \Delta_{vc\mathbf{k}}(\mathbf{q},\omega) M^{\mathbf{G}'}_{vc\mathbf{k}}(\mathbf{q})
% \end{align}

\section{Computational Details}
All calculations presented here were performed on Perlmutter (NERSC), Kestrel (NREL), and Frontier (OLCF) supercomputers. 
For the NERSC Perlmutter computer system a GPU node is composed of 4 NVIDIA A100 GPUs with a single AMD EPYC 7763 (Milan) CPU. Calculations on Perlmutter GPU nodes were run with 4 MPI ranks per node and 16 OpenMP threads per rank. 
A CPU node on Perlmutter is composed of 2 AMD EPYC 7763 (Milan) CPUs, and calculations on Perlmutter CPU nodes used 16 MPI ranks per node and 8 OpenMP threads per rank. 
The NREL Kestrel computer system is composed of GPU nodes containing 4 NVIDIA H100 GPUs and 2 AMD EPYC 9554 (Genoa) CPUs. Calculations on Kestrel GPU nodes were run with 4 MPI ranks per node and 16 OpenMP threads per rank.
The OLCF Frontier GPU node is composed of 4 AMD Instinct MI250X GPUs, each comprised of 2 Graphics Compute Dies (GCD) for a total of 8 devices with a single AMD EPYC 7763 (Milan) CPU. Calculations on Frontier GPU nodes were run with 8 MPI ranks per node (one rank per GCD) and 7 OpenMP threads per rank.

DFT calculations for the Pt(111) and Pt(111)+H surfaces used for scaling tests were performed with the Quantum ESPRESSO code~\cite{Giannozzi_2009,Giannozzi_2017} using an 80 Ryd plane-wave cutoff, SG15 ONCV norm-conserving pseudopotentials,~\cite{SCHLIPF201536} and the Perdew-Burke-Ernzerhof (PBE) generalized gradient approximation (GGA) functional.~\cite{PhysRevLett.77.3865} All calculations used 1st order Methfessel-Paxton smearing with a broadening of 0.2 eV.\cite{methfessel_high-precision_1989} A 12$\times$12$\times$12 \textbf{k}-point grid was used for the bulk primitive fcc Pt cell lattice relaxation. The DFT-predicted Pt lattice constant was 3.976 Å, which is in good agreement with the experimental value of 3.924 Å and excellent agreement with the values previously reported for all-electron (3.971 Å) and SG15 pseudopotential (3.975 Å) calculations.\cite{haynes2014crc} To create the 4-layer (1$\times$1) Pt(111) surface, the bulk Pt cell was expanded into a 1$\times$1$\times$4 supercell, the lattice vectors parallel to the Pt surface were held constant, and 20 Å of vacuum space was added. Ionic optimization was performed on the clean and 1 ML H surfaces with the lower two layers of the slab frozen. The larger Pt(111) or Pt(111)+H supercells were created by expanding the (1$\times$1) Pt(111) or Pt(111)+H surfaces, respectively, along the \textbf{a} and \textbf{b} lattice vectors without further ionic optimization. The \textbf{k}-point grid for these surfaces was reduced proportionally to the increase in surface area. A summary of the chosen \textbf{k}-point grids is shown in Table 1. For the surfaces, Coulomb-truncation was applied along the z axis, which was perpendicular to the surface.~\cite{wignerseitz} Empty state generation was performed using BerkeleyGW. 

The (4$\times$4) Pt(111) surface with explicit solvation and Nafion fragment was created using the JDFTx software package~\cite{SUNDARARAMAN2017278} but otherwise the same DFT settings and surface generation procedure as above. The DFT-predicted Pt lattice constant was still 3.976 Å. An O$_2$ molecule bidentate-bound across two Pt atoms, an unadsorbed H$_3$O molecule, and an unadsorbed deprotonated representative Nafion fragment with the molecular formula CF$_3$OCF$_2$CF$_2$SO$_3$ were then added to the system. The volume to be filled with explicit water molecules was calculated by subtracting the volume contained by the isosurface containing 99\% of the electrons of the Pt(111)+O$_2$$^*$+H$_3$O+Nafion system from the total cell volume. 60 water molecules were added so that the water density in this region was closest to the experimental value of 0.997 g/mL at 1 atm and 300 K.\cite{NIST} This system was equilibrated using AIMD in the NPT ensemble for 500 fs with 1 fs timesteps at 300 K and 1 bar along the z axis using JDFTx, a 2$\times$2$\times$1 Monkhorst-Pack \textbf{k}-grid, and otherwise the same settings as above. A final DFT calculation without ionic optimization using the final equilibrated structure and same k-grid was used to generate the DFT charge density and wavefunctions for the subsequent RPA jobs. The O$_2$ molecule from this equilibrated structure was deleted, with all other atoms fixed in place, to create the reference surface for the O$_2$ adsorption calculation. The Nafion, H$_2$O, and H$_3$O were deleted from the equilibrated AIMD structure to create a (4$\times$4) Pt(111)+O$_2$ structure without solvation, and the Pt and O atoms were again not allowed to relax. Although outside the scope of the current work, future studies should be done to fully quantify the impact of configurational sampling in these systems. The triplet O$_2$ molecule reference energy was calculated using the above settings and a box size of 10.0 $\times$ 10.1 $\times$ 10.2 $\angstrom$. Empty state generation was performed using JDFTx.

All RPA scaling calculations were run using BerkeleyGW with screened cutoffs of 10, 20, and 30 Ryd in the full-frequency Adler-Wiser formalism with the same \textbf{k}/\textbf{q}-grid as in the DFT calculations. The number of bands used in each screened cutoff calculation was the maximum number of \textbf{G}-vectors with kinetic energy less than the screened cutoff. A Gauss-Legendre quadrature with 16 points was used for the frequency integration along the imaginary axis.\cite{rALDA_npj} The RPA calculations used the same partial occupations settings as the DFT calculations. The static subspace approximation using 40\% of the total basis set was used. The (4$\times$4) Pt(111)+Nafion+H$_2$O+H$_3$O+O$_2$ calculations were run using screened cutoffs of 17, 18, 19, 20, 21, and 22 Ryd to extrapolate to infinite cutoff and 12 imaginary frequencies, but otherwise the same settings as the scaling tests. We show in another work that these settings for the subspace approximation and screened cutoffs produce robust RPA correlation energies. 

All exact exchange calculations were performed using JDFTx using Wigner-Seitz exchange regularization and the same grid as the DFT calculations. 

\begin{table}[th]
\caption{Summary of the \textbf{k}/\textbf{q}-point grids used for the different Pt(111) surfaces. }
\label{tab:ptSurfaces}
\begin{tabular}{cc}
\begin{tabular}[c]{@{}c@{}}Pt(111) supercell\\ expansion along (\textbf{a}, \textbf{b})\end{tabular} & \textbf{k}/\textbf{q}-point grid \\ \hline
1$\times$1 & 12$\times$12$\times$1 \\
2$\times$1 & 6$\times$12$\times$1 \\
2$\times$2 & 6$\times$6$\times$1 \\
3$\times$2 & 4$\times$6$\times$1 \\
3$\times$3 & 4$\times$4$\times$1 \\
4$\times$3 & 3$\times$4$\times$1 \\
4$\times$4 & 3$\times$3$\times$1 \\
\hline
\end{tabular}
\end{table}

\section{Performance Results}

\begin{figure*}[th!]
    \centering
    \includegraphics[width=\textwidth]{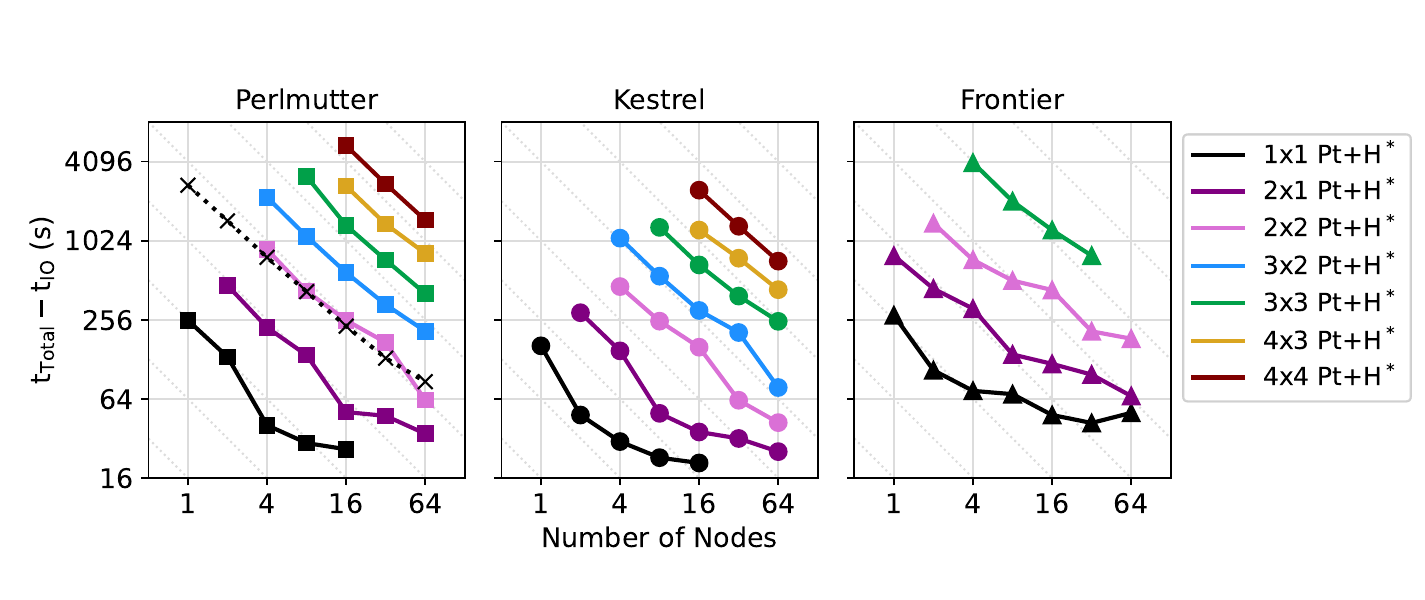}
    \caption{Strong Scaling for (left) Perlmutter GPU (solid lines) and CPU (dotted line), (center) kestrel GPU and (right) Frontier, shows how the total computational time (in seconds), subtracting I/O time, changes as the number of compute nodes is increased.}
    \label{fig:strongScale}
\end{figure*}

% By alleviating the memory bottleneck in the \texttt{CHI\_SUM} kernel and moving most of the computational work onto the GPU, a nearly 10-fold speedup is achieved over the CPU-only implementation. \textcolor{red}{Feel like I should instead compare this to the old algorithm on GPUs}
By alleviating the memory bottleneck in the \texttt{CHI\_SUM} kernel and moving much of the computational work onto the GPU, the \texttt{CHI\_SUM} kernel achieves a speedup by a factor of two over the previous implementation. Additionally the calculation can be carried out on a smaller number of nodes without as severe of a requirement on the total available memory.
%It looks to me that this doesn't actually have different scaling. the wavefuctions grow as N and actually M also grows as N. 

We calculated the RPA energy for a series of differently sized hydrogenated platinum surfaces, detailed in Table \ref{tab:ptSurfaces}. These surfaces ranged from 4 Pt atoms to 64 Pt atoms and included 20 Å of vacuum along the z-direction, perpendicular to the surface. For each of these systems, the RPA energy was calculated at a single $\mathbf{q}$-point using either 10, 20, or 30 Rydberg screening cutoffs along a frequency integration quadrature of 16 frequencies. Depending on the cutoff used, a different number of conduction states were considered, chosen such that the total number of bands included was equal to the number of $\mathbf{G}$ vectors within that screened cutoff. As seen in Figure \ref{fig:sizePerf}, across this wide range range of system sizes the number of bands in the calculation is the key determining factor of the computational cost. Additionally the relationship between the computational cost for a single $\mathbf{q}$-point and the number of bands in the calculation follows a quadratic scaling across almost the entire domain of interest, only deviating slightly towards cubic scaling as the number of bands approaches 40 thousand. 

This phenomenological performance can be understood by considering the weak scaling shown in Figure \ref{fig:weakScale}. Since the number of $k$-points, $N_k$, has an inverse relationship to the system size and all of $N_v$, $N_c$, and $N_\psi$ are linearly related to system size the \texttt{MTXEL} kernel should show a quadratic dependence on system size. That is exactly what is seen in Figure \ref{fig:weakScale}. On the other hand, the \texttt{CHI\_SUM} kernel is expected to show cubic dependence on system size. For the systems studied here, however, the overall computational cost is driven by the \texttt{MTXEL} kernel. This is both due to the large vacuum region included to converge the calculation and the efficiency of the optimized \texttt{CHI\_SUM} kernel. For the largest systems studied here, the \texttt{CHI\_SUM} kernel rivals the \texttt{MTXEL} kernel in computational expense, leading to the onset of cubic scaling. 

It is worth repeating that these results are for a single $\mathbf{q}$-point, and that the number of $\mathbf{q}$-points needed to maintain the same reciprocal space grid spacing is also inversely related to the system size. This means that the quadratic and cubic scaling of the \texttt{MTXEL} and \texttt{CHI\_SUM} kernels, respectively, for a single $\mathbf{q}$-point should be expected to revert to linear and quadratic scaling for the case of the full $\mathbf{q}$-point sampling. This means that, for systems of scientific interest containing up to 40 thousand bands, full RPA correlation energies can be obtained at a computational cost only linearly dependent on the system size.

% \begin{figure}
%     \centering
%     \includegraphics[width=\columnwidth]{implementation_figures/StrongScaling.pdf}
%     \caption{Strong Scaling for Perlmutter GPU (dashed lines), Perlmutter CPU (dotted line), and Frontier (dot-dashed lines).}
%     \label{fig:strongScale}
% \end{figure}

This algorithm scales extremely well to hundreds of GPUs and is implemented in a portable manner that can achieve strong performance across HPC systems. In Figure \ref{fig:strongScale} the strong scaling of the code is shown for a wide range of calculation sizes across several HPC systems. This shows the near-ideal scaling of the code performance with respect to the number of compute nodes used. 

The Perlmutter GPU nodes show over a 10 times speedup over the CPU nodes, and the Kestrel GPU nodes are around twice as fast as the Perlmutter GPU nodes. Since the H100 chips on the Kestrel GPU nodes have 80GB of on-chip memory compared to the 40GB on most of the Perlmutter A100 chips, fewer valence blocks are needed on Kestrel as compared to Perlmutter. 

One notable feature of Figure \ref{fig:strongScale} is that there are cases, such as going from 1 to 2 Kestrel nodes (or from 2 to 4 Perlmutter GPU nodes) for the Pt+H* 1x1 and going from 4 to 8 Kestrel nodes on the 2x1, where the observed scaling relationship is better than the "ideal" inverse relationship. This occurs because at that extent of parallelization there is only a single valence block and the \texttt{MTXEL} kernel only needs to run a single time. This results in an additional computational saving and the apparent "better than ideal" scaling.

\subsection{Application towards heterogenous interfaces}
Ab-initio studies of complex heterogenous interfaces will become more important as advanced (electro)catalysts, electrochemical devices, and battery devices are made. The algorithmic implementation discussed in this work enables beyond-DFT methods to be more computationally accessible as a potential option in the study of these systems. As a demonstration of chemical systems possible, we calculated the adsorption energy of an O$_2$ molecule to a (4$\times$4) Pt(111) surface interfaced with water and a small Nafion fragment using both DFT and RPA, which is an important interface in fuel cell devices and the oxygen reduction reaction. We note that since the number of bands used in the calculation depends on the plane-wave/screened cutoffs used and system volume, the extra atoms included due to solvation are not the dominant factor in determining the cost of the RPA correlation energy calculation. However, additional work is still needed to efficiently explore the vast configurational space possible in this explicitly solvated system and account for the how the solvent responds to the presence of a solute in the calculation of an adsorption energy. We emphasize that this calculation is meant to showcase the capability. 

\begin{figure}
   \centering
    \includegraphics[width=0.35\linewidth]{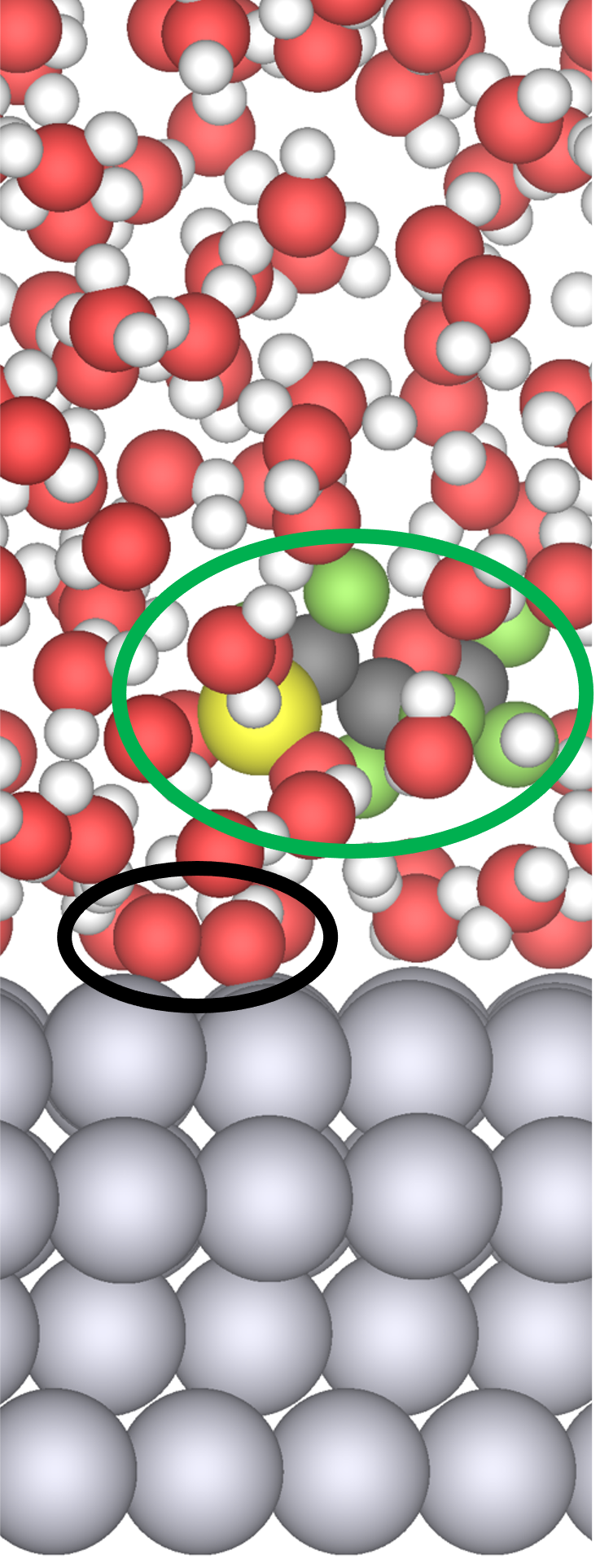}
    \caption{Structure of the Pt(111)+Nafion+60 H$_2$O+H$_3$O+O$_2$$^*$ fragment system after AIMD equilibration in the NPT ensemble at 300 K and 1 bar after 500 fs. Light gray, red, white, yellow, dark gray, and green spheres represent Pt, O, H, S, C, and F atoms, respectively. The black circle highlights the O$_2$ molecule adsorbed to the surface in a bidentate configuration. The green circle highlights the Nafion fragment.}
    \label{fig:heroSystem}
\end{figure}

Figure \ref{fig:heroSystem} shows the structure used to calculate the adsorption energy and Table \ref{tab:heroEads} summarizes the adsorption energies for this system configuration using DFT, DFT with the Grimme D3 van der Waals correction,~\cite{10.1063/1.3382344,https://doi.org/10.1002/jcc.21759} and RPA with exact exchange. First, the O$_2$ adsorption energy is destabilized with RPA as compared to DFT in both the vacuum (0.17 eV) and solvated cases (0.08 eV). The increase in adsorption energy using RPA as compared to DFT has been previously explained for CO on Pt(111) as caused by the DFT underestimation of the frontier orbital energy gap of the adsorbing CO.~\cite{Schimka2010} Thus, we propose that similar behavior occurs here. Next, it is apparent that the solvated environment significantly decreases the adsorption energy of O$_2$ to Pt as compared to the vacuum case, which is expected given the additional interactions of O$_2$ with Nafion and water. This decrease is predicted to be -1.12, -1.30, and -1.21 eV for DFT, DFT+D3, and RPA, respectively. Overall, the relatively small shifts in O$_2$ adsorption energy with RPA suggests beyond-DFT may not play as important of a role in this system as compared to other possible (photo)electrochemical systems involving more localized d-electrons, such as TiO$_2$ or MoS$_2$. 

\renewcommand{\arraystretch}{1.3}

\begin{table}[]
\caption{Summary of O$_2$ adsorption energies in eV for the vacuum and solvated (4$\times$4) Pt(111) surfaces. The $\Delta$ column and row show the change in adsorption energy due to the presence of the Nafion+60 H$_2$O+H$_3$O and functional used, respectively.}
\label{tab:heroEads}
\begin{tabular}{c|ccc}
                       & Pt(111)     & \begin{tabular}[c]{@{}c@{}}Pt(111)+Nafion+\\ 60 H$_2$O+H$_3$O\end{tabular} & $\Delta_{solvent}$  \\ \hline
E$_{ads,O_2}^{DFT}$    & -0.342 & -1.462                                                                & -1.120 \\
E$_{ads,O_2}^{DFT+D3}$ & -0.583 & -1.882                                                                & -1.299 \\
E$_{ads,O_2}^{RPA}$    & -0.176 & -1.384                                                                & -1.209 \\
$\Delta_{RPA-DFT}$     & 0.166  & 0.078                                                                 &       
\end{tabular}
\end{table}

% \begin{outline}[enumerate]
% %\1 Computational cost with increasing E\_cut
% \1 Hero Calc?

% \end{outline}

\section{Conclusion}
In conclusion, we present an efficient and highly scalable code to calculate the RPA correlation energies of large systems. We have identified the number of bands in the calculation as the key parameter controlling computational cost for the surface catalytic systems under consideration here. We have further shown that the computational cost of a RPA energy calculation for these systems will scale linearly with the number of bands up to around 40 thousand bands after which point quadratic scaling will dominate. 
We also note in passing that recent work has shown that the computational cost and scaling with system size can be further reduced using a "stochastic pseudobands" compression of the input wavefunctions, and that this approach is fully compatible with the implementation described herein.\cite{PhysRevLett.132.086401}

The performance reported in this work was achieved through an extensive parallelization scheme and the implementation of a batched algorithm for the summation over conduction-valence pairs on GPUs that avoids memory or communication bottlenecks. This implementation is portable across leadership-class HPC systems and will significantly accelerate the generation of high-quality datasets crucial to furthering our understanding of key catalytic systems. 

\section{Acknowledgments}
This work was primarily supported by the Beyond-DFT Electrochemistry with Accelerated and Solvated Techniques (BEAST) project as part of the Computational Chemical Sciences program, funded by the U.S. Department of Energy, Office of Science, Basic Energy Sciences, award No. DE-SC0022247.
M.D.B. acknowledges support by the Center for Computational Study of Excited-State Phenomena in Energy Materials (C2SEPEM), under contract No. DE-AC02-05CH11231.
We used computational resources sponsored by the Department of Energy’s Office of Energy Efficiency and Renewable Energy, located at the National Renewable Energy Laboratory; Oak Ridge Leadership Computing Facility at the Oak Ridge National Laboratory, which is supported by the Office of Science of the U.S. Department of Energy under Contract No. DE-AC05-00OR22725; and at the National Energy Research Scientific Computing Center (NERSC), a U.S. Department of Energy Office of Science User Facility located at Lawrence Berkeley National Laboratory, operated under Contract No. DE-AC02-05CH11231 using NERSC award ERCAP-m4025.

\bibliography{bibliography}

\end{document}